\documentclass[twocolumn]{aa}

\usepackage{graphicx}
\usepackage{amsmath,amsfonts,amssymb}
\usepackage{txfonts}
\usepackage{color}
\usepackage{natbib}
\usepackage{float}
\usepackage{dblfloatfix}
\usepackage{afterpage}
\usepackage{ifthen}
\usepackage[morefloats=12]{morefloats}
\usepackage{placeins}
\usepackage{multicol}
\bibpunct{(}{)}{;}{a}{}{,}
\usepackage[switch]{lineno}
\definecolor{linkcolor}{rgb}{0.6,0,0}
\definecolor{citecolor}{rgb}{0,0,0.75}
\definecolor{urlcolor}{rgb}{0.12,0.46,0.7}
\usepackage[breaklinks, colorlinks, urlcolor=urlcolor,
    linkcolor=linkcolor,citecolor=citecolor,pdfencoding=auto]{hyperref}
\hypersetup{linktocpage}
\usepackage{bold-extra}
\usepackage{csquotes}
\usepackage{breakurl}

\usepackage{upgreek}

\newcommand{\ve}[1]{{\vec #1}}
\newcommand{\ma}[1]{\tens{#1}}
\newcommand{\pcal}{{P}}

\newcommand{\BP}{\textsc{BeyondPlanck}}

\newcommand{\Madam}{\texttt{Madam}}
\newcommand{\e}{\mathrm{e}}
\def\WMAP{\emph{WMAP}}

\def\Planck{\textit{Planck}}

\def\commander{\texttt{Commander}}

\def\arcm{\ifmmode {^{\scriptstyle\prime}}
          \else $^{\scriptstyle\prime}$\fi}
\providecommand{\sorthelp}[1]{}          

\begin{document}

\title{\bfseries{\scshape{BeyondPlanck}} II. CMB map-making through Gibbs sampling }
	\newcommand{\nersc}[0]{1}
\newcommand{\princeton}[0]{2}
\newcommand{\helsinkiA}[0]{3}
\newcommand{\milanoA}[0]{4}
\newcommand{\triesteA}[0]{5}
\newcommand{\haverford}[0]{6}
\newcommand{\helsinkiB}[0]{7}
\newcommand{\triesteB}[0]{8}
\newcommand{\milanoB}[0]{9}
\newcommand{\milanoC}[0]{10}
\newcommand{\oslo}[0]{11}
\newcommand{\jpl}[0]{12}
\newcommand{\mpa}[0]{13}
\newcommand{\planetek}[0]{14}
\author{\small
E.~Keih\"{a}nen\inst{\helsinkiA}\thanks{Corresponding author: E.~Keih\"anen; \url{elina.keihanen@helsinki.fi}}
\and
A.-S.~Suur-Uski\inst{\helsinkiA,\helsinkiB}
\and
K.~J.~Andersen\inst{\oslo}
\and
R.~Aurlien\inst{\oslo}
\and
R.~Banerji\inst{\oslo}
\and
M.~Bersanelli\inst{\milanoA,\milanoB,\milanoC}
\and
S.~Bertocco\inst{\triesteB}
\and
M.~Brilenkov\inst{\oslo}
\and
M.~Carbone\inst{\planetek}
\and
L.~P.~L.~Colombo\inst{\milanoA}
\and
H.~K.~Eriksen\inst{\oslo}
\and
M.~K.~Foss\inst{\oslo}
\and
C.~Franceschet\inst{\milanoA,\milanoC}
\and
U.~Fuskeland\inst{\oslo}
\and
S.~Galeotta\inst{\triesteB}
\and
M.~Galloway\inst{\oslo}
\and
S.~Gerakakis\inst{\planetek}
\and
E.~Gjerl{\o}w\inst{\oslo}
\and
B.~Hensley\inst{\princeton}
\and
D.~Herman\inst{\oslo}
\and
M.~Iacobellis\inst{\planetek}
\and
M.~Ieronymaki\inst{\planetek}
\and
H.~T.~Ihle\inst{\oslo}
\and
J.~B.~Jewell\inst{\oslo}
\and
A.~Karakci\inst{\oslo}
\and
R.~Keskitalo\inst{\nersc}
\and
G.~Maggio\inst{\triesteB}
\and
D.~Maino\inst{\milanoA, \milanoB, \milanoC}
\and
M.~Maris\inst{\triesteB}
\and
A.~Mennella\inst{\milanoA,\milanoB,\milanoC}
\and
S.~Paradiso\inst{\milanoA,\milanoC}
\and
B.~Partridge\inst{\haverford}
\and
M.~Reinecke\inst{\mpa}
\and
T.~L.~Svalheim\inst{\oslo}
\and
D.~Tavagnacco\inst{\triesteB, \triesteA}
\and
H.~Thommesen\inst{\oslo}
\and
M.~Tomasi\inst{\milanoA, \milanoB}
\and
D.~J.~Watts\inst{\oslo}
\and
I.~K.~Wehus\inst{\oslo}
\and
A.~Zacchei\inst{\triesteB}
}
\institute{\small
Computational Cosmology Center, Lawrence Berkeley National Laboratory, Berkeley, California, U.S.A.\goodbreak
\and
Department of Astrophysical Sciences, Princeton University, Princeton, NJ 08544,
U.S.A.\goodbreak
\and
Department of Physics, Gustaf H\"{a}llstr\"{o}min katu 2, University of Helsinki, Helsinki, Finland\goodbreak
\and
Dipartimento di Fisica, Universit\`{a} degli Studi di Milano, Via Celoria, 16, Milano, Italy\goodbreak
\and
Dipartimento di Fisica, Universit\`{a} degli Studi di Trieste, via A. Valerio 2, Trieste, Italy\goodbreak
\and
Haverford College Astronomy Department, 370 Lancaster Avenue,
Haverford, Pennsylvania, U.S.A.\goodbreak
\and
Helsinki Institute of Physics, Gustaf H\"{a}llstr\"{o}min katu 2, University of Helsinki, Helsinki, Finland\goodbreak
\and
INAF - Osservatorio Astronomico di Trieste, Via G.B. Tiepolo 11, Trieste, Italy\goodbreak
\and
INAF/IASF Milano, Via E. Bassini 15, Milano, Italy\goodbreak
\and
INFN, Sezione di Milano, Via Celoria 16, Milano, Italy\goodbreak
\and
Institute of Theoretical Astrophysics, University of Oslo, Blindern, Oslo, Norway\goodbreak
\and
Jet Propulsion Laboratory, California Institute of Technology, 4800 Oak Grove Drive, Pasadena, California, U.S.A.\goodbreak
\and
Max-Planck-Institut f\"{u}r Astrophysik, Karl-Schwarzschild-Str. 1, 85741 Garching, Germany\goodbreak
\and
Planetek Hellas, Leoforos Kifisias 44, Marousi 151 25, Greece\goodbreak
}

	\authorrunning{}
	\titlerunning{Map-making by Gibbs sampling}

\abstract{We present a Gibbs sampling solution to the map-making
  problem for CMB measurements, building on existing destriping
  methodology.  Gibbs sampling breaks the computationally heavy
  destriping problem into two separate steps; noise filtering and map
  binning.  Considered as two separate steps, both are
  computationally much cheaper than solving the combined problem.
  This provides a huge performance benefit as compared to traditional
  methods, and allows us for the first time to bring the destriping
  baseline length to a single sample.  We apply the
  Gibbs procedure to simulated \textit{Planck} 30 GHz data.  We find that
  gaps in the time-ordered data are handled efficiently by filling
  them with simulated noise as part of the Gibbs process.  The Gibbs
  procedure yields a chain of map samples, from which we may compute
  the posterior mean as a best-estimate map.  The variation in the
  chain provides information on the correlated residual noise, without
  need to construct a full noise covariance matrix. However, if only a
  single maximum-likelihood frequency map estimate is required, we
  find that traditional conjugate gradient solvers converge much
  faster than a Gibbs sampler in terms of total number of
  iterations. The conceptual advantages of the Gibbs sampling approach
  lies in statistically well-defined error propagation and systematic
  error correction, and this methodology forms the conceptual basis
  for the map-making algorithm employed in the \textsc{BeyondPlanck} framework, which
  implements the first end-to-end Bayesian analysis pipeline for CMB
  observations.  }


\maketitle

\section{Introduction}

Removal of correlated $1/f$ noise generated by detectors is a crucial
step in the data processing chain for cosmic microwave background
(CMB) measurements.  Noise removal is most commonly done jointly with
map-making.  A number of different methods have been developed for
this purpose.  For reviews of different methods applied to \Planck,
see \citet{Poutanen2006},
\citet{Ashdown2007a,Ashdown2007b,Ashdown2009}, and references
therein. For a closely related discussion of Bayesian noise estimation
with time-ordered CMB data, see \citet{wehus:2012}.

Conventional map-making produces as initial output pixelized sky maps
of CMB temperature and polarization at a given frequency.  These
pixelized sky maps then serve as input for the separation of
astrophysical components \citep[e.g.,][]{planck2016-l04}, including
the CMB, and further for power spectrum estimation
\citep[e.g.,][]{planck2016-l05}.  As input, map-making takes the
calibrated time-ordered information (TOI) data stream, together with
corresponding detector pointing information.

The generalized least-squares methods (GLS) aim at finding the map
$\ve m$ that maximizes the likelihood of the data ${P}(\ve d \mid \ve
m)$, where the likelihood depends on an assumed known noise spectrum.
Another widely used approach is the destriping technique
\citep{Burigana1997,delabrouille1998a, maino1999,maino2002b,
  keihanen2004, keihanen2005,Sutton09,kurki-suonio2009,keihanen2010}, where the
correlated noise component is modelled as a sequence of offsets, whose
amplitudes are then solved for through maximum likelihood analysis and
then subtracted from the timeline.

Conventional map-making, whether GLS or destriping, is a
memory-intensive data processing step, since it requires that all the
detector pointing information for the data set is kept in memory
simultaneously.  This follows from the coupling between two signal
components with very different characteristics: the sky signal
(represented by a map) is dependent on the detector's pointing on the
sky, while the noise component is correlated in the time domain but
independent of where the detector is pointing.

In this work we examine the possibility of solving the map-making and
noise removal problem through a Gibbs sampling technique.  Gibbs
sampling breaks the computationally heavy map-making problem into two separate steps:
noise removal and construction of the sky signal from noise-cleaned
TOI.  Considered as isolated steps, both are much simpler than the
combined map-making procedure.  As a test case we use simulated \Planck\ LFI
data in the context of the \BP\ project.  For an introduction to Gibbs
sampling theory, and for an overview of the \BP\ project, we refer the
reader to the first paper of this series, \cite{bp01}.

A great benefit of the proposed procedure is that the noise removal
step can be carried out separately for each \Planck\ pointing period,
reducing the memory requirement tremendously, as compared to
conventional map-making methods. Second, in the absence of data
flagging and Galactic masking, the noise removal step reduces into a
simple Fast Fourier Transform (FFT) filtering operation, in which case
the computational cost of noise removal is equivalent to only two FFTs
of the full TOI.  Flagging, however, breaks the stationarity of the
data, and we examine two alternative solutions to this problem:
filling of gaps with a Gibbs technique, and exactly solving the
non-stationary filtering problem.  As a consequence of these two
facts, we are for the first time able to reduce the length of the
destriping baseline to one single TOI sample, with a significantly
lower combined computational cost than traditional approaches.

As a byproduct, the Gibbs map-making process also yields an estimate
of the residual noise in the output products in the form of a discrete
set of samples that accounts for both white and residual correlated
noise, without need to compute a noise covariance matrix. The
computational cost of running a full Gibbs chain, like that described
in this paper, must therefore be compared to the cost of generating a
full ensemble of end-to-end simulations, or the cost of evaluating a
dense noise covariance matrix, in a traditional pipeline.

The main focus of this paper is on the noise removal step.  For
demonstration purposes, we combine noise removal with a simple
pixel-based map-making procedure.  For this purpose we have written a
stand-alone test code, which allows us to study the map-making step in
separation, independently from the \commander\ code
\citep{eriksen2008} that forms the basis for the \BP\ processing
\citep{bp01}.  The great potential of the method, however, lies in a
scenario where noise removal is combined within the Gibbs framework
with modeling of the sky signal and instrument effects, which is the
main overall goal of the \BP\ project.

\section{Methodology}
\label{sec:method}

\subsection{Gibbs sampling procedure}

We consider a time-ordered data stream from one detector
of a \Planck-like CMB experiment, which may be modelled as
\begin{equation}
  \ve{d} = \ma{P}\ve{m}  +\ve{n'}.
\end{equation}
Here $\ma{P}$ is the pointing matrix, which encodes the scanning
strategy and the detector's response to temperature and polarization,
and $\ve{m}$ is the pixelized sky map, which includes temperature and
polarization components in the form of $I,Q,U$ Stokes components.
Formally, $\ma P$ is a large matrix of size $(N_t,3N_p)$, where $N_t$
is the number of samples in the time-ordered data stream, and $N_p$ is
the number of pixels in the sky map.  In the place of $\ma{P}\ve{m}$
we could have other sky models, for instance a harmonic representation
of the sky, or a parametrized foreground model.  In this work we do
not pursue these possibilities further, but assume the simple
pixelized sky model.

The instrument noise of \Planck\ LFI radiometers is 
well approximated as Gaussian distributed \citep{Bersanelli2010}. 
The noise stream can be divided into two components: correlated $1/f$ noise,
and uncorrelated white noise.
In the spirit of destriping, we model the correlated noise component
as a sequence of constant offsets, or ``baselines'' of fixed length $N_a$.
We write the noise term as
\begin{equation}
  \ve{n'} = \ma{F}\ve{a}  +\ve{n},
\end{equation}
where $\ve{a}$ represents the baseline amplitudes, and
$\ma{F}$ is a matrix which formally projects them into a full data
stream.  With a baseline length of $N_a$, each column of $\ma{F}$ has
the value 1 along $N_a$ adjacent elements, and the value 0 elsewhere.
In the extreme limit in which the baseline length equals one sample,
$\ma{F}$ becomes a unity matrix and can be dropped from the equations.
The last term, $\ve{n}$, represents white noise. 

We denote the
covariance matrices of $\ve a$ and $\ve n$ by $\ma C_a$ and $\ma C_w$, respectively, where $\ma
C_w$ is diagonal (but not necessarily uniform), and $\ma C_a$ is the
covariance of the noise baseline amplitudes.  If the noise is
stationary, the latter is band-diagonal to a good approximation, and
can be represented as a filter in Fourier domain. If the baseline length
equals one TOI sample ($N_a=1$), $\ma C_a$ becomes equal to the
time-domain covariance of the $1/f$ noise component.  The construction
of the covariance in the general case of $N_a>1$ is presented in
\cite{keihanen2010}; in this work we will assume $N_a=1$.

In the following, both covariance matrices are assumed to be known
a priori.  Estimating the noise properties from the data itself
(as part of the Gibbs sampling process) is addressed by \citet{bp08}.

We now proceed to write out the posterior distribution, $\pcal(\ve
m,\ve a\mid \ve y, \ma C_w,\ma C_a)$, for the $(\ve a,\ve m)$ model,
given some TOI data stream $\ve y$ and assumed known noise properties.
This is the posterior distribution we eventually want to sample by
using Gibbs sampling technique, and may, according to Bayes' theorem,
be written as 
\begin{equation}
\pcal(\ve m,\ve a\mid \ve y, \ma C_w,\ma C_a) = 
\pcal(\ve y \mid \ve m,\ve a,\ma C_w) \frac{\pcal(\ve m)\pcal(\ve a \mid \ma C_a)}{\pcal (\ve y)}. 
\end{equation}
Here we have assumed that $\ve a$ are $\ve m$ are statistically
independent, and we include $\ma C_a$ and $\ma C_w$ explicitly only in
factors for which the conditional in question actually depends on
them.

The denominator $\pcal(\ve y)$ is an overall normalization factor and
can be ignored, since the methods we use for drawing samples from the
likelihood are insensitive to the normalization.  Furthermore, we assume a
uniform prior for the sky map $\ve m$, $\pcal(\ve m)=1$, such that
\begin{equation}
\pcal(\ve m,\ve a\mid \ve y, \ma C_w,\ma C_a) \propto
\pcal(\ve y \mid \ve m,\ve a, \ma C_w) \pcal(\ve a \mid \ma C_a). \label{like1}
\end{equation}
The first factor on the right is the probability of observing a data
stream $\ve y$ for given realization of correlated noise $\ve a$ and
for a given sky map $\ve m$.  With both $\ve a$ and $\ve m$ fixed, the
only difference between the model and the data comes from white noise.
We obtain this likelihood from the white noise distribution, which
is assumed to be Gaussian with covariance $\ma C_{w}$,
\begin{equation}
\pcal(\ve y \mid \ve m,\ve a, \ma C_w) =
\frac{\e^{-\frac12(\ve y - \ma{F}\ve{a} -\ma{P}\ve{m} )^T\ma{C}_w^{-1}
                 (\ve y - \ma{F}\ve{a} -\ma{P}\ve{m} ) }}{\sqrt{|2\pi\ma{C}_w|}}.
\end{equation}
The second distribution on the right-hand side of Eq.~(\ref{like1})
represents a prior on $\ve a$, i.e., the probablility of obtaining a
given realization of correlated noise in the abscence of actual
measurements.  As is customary in \Planck\ map-making, we assume that
the correlated noise component also is Gaussian distributed, such that
\begin{equation}
\pcal(\ve a \mid \ma C_a) = \frac{\e^{-\ve{a}^T\ma{C}_a^{-1}\ve
    a}}{\sqrt{|2\pi\ma{C}_a|}}.
\end{equation}

The conventional destriping procedure finds the combination $(\ve
a,\ve m)$ that maximizes the posterior distribution in
Eq.~(\ref{like1}).  Equating the derivative of this expression with
respect to $(\ve a,\ve m)$ to zero leads to a large linear system,
which can be solved by conjugate gradient iteration \citep[see, e.g.,][]{keihanen2010}.

In this paper, we instead proceed to use Gibbs sampling to sample
from the distribution in Eq.~(\ref{like1}).  One full Gibbs sampling
step consists of two substeps, in each of which one of the two
parameters $\ve m,\ve a$ is kept fixed, and the other is drawn from
the corresponding conditional distribution,
\begin{eqnarray}
 \ve{m'}  &\leftarrow& \pcal(\ve m \mid \ve a;,\ve{y},\ma C_w)   \nonumber \\
 \ve{a'}  &\leftarrow& \pcal(\ve a \mid \ve m;\ve{y},\ma C_w,\ma C_a)  
\end{eqnarray}
We use the semicolon to separated sampling parameters, which are however 
considered constant in the current sampling step,
from parameters not sampled.
Here the symbol ``$\leftarrow$'' indicates drawing a sample from the
distribution on the right-hand side. 
Let us now look at these two
substeps more closely:
\begin{enumerate}
\item For a given data stream $\ve y$ and noise baselines $\ve a$,
    we update the sky map $\ve m$ by drawing a sample
    from the Gaussian distribution
\begin{equation}
\pcal(\ve m \mid \ve a;\ve y,\ve C_w) \propto
\frac{\e^{-\frac12(\ve y' -\ma{P}\ve{m} )^T\ma{C}_w^{-1}
                 (\ve y' -\ma{P}\ve{m} ) }}{\sqrt{|2\pi\ma{C}_w|}}  \label{slice1}
\end{equation}
where $\ve y'=\ve y-\ma F\ve a$ is the data stream
from which we have subtracted
the current $\ve a$ sample. 
Thus $\ve y'$ represents the current estimate of the noise-cleaned TOI stream.
\item For given data stream $\ve y$ and sky $\ve m$,
    we update the baseline vector $\ve a'$
    by drawing a sample
from the distribution
\begin{equation}
\pcal(\ve a \mid \ve m;\ve y,\ma C_w,\ma C_a) \propto \frac{\e^{-\frac12(\ve y'' -\ma{F}\ve{a} )^T\ma{C}_w^{-1}
                 (\ve y'' -\ma{F}\ve{a} ) }}{\sqrt{|2\pi\ma{C}_w|}}
                 \frac{\e^{-\ve{a}^T\ma{C}_a^{-1}\ve a}}{\sqrt{|2\pi\ma{C}_a|}}, \label{slice2}
\end{equation}
where now $\ve y''=\ve y-\ma P\ve m$
is a data stream from which we have subtracted 
the current $\ve m$ sample. Thus $\ve y''$ represents the current estimate
of the noise-only TOI stream.
\end{enumerate}
The outcome from the sampling procedure is a chain
of $\ve m$ and $\ve a$ samples,
with their distribution sampling the combined likelihood
of Eq.~(\ref{like1}).

The conditional distributions in Eqs.~(\ref{slice1}) and (\ref{slice2})
are both multivariate Gaussian distributions, from which samples may
be drawn efficiently using standard methods; see, e.g., Appendix A in
\citet{bp01} and references therein. In the following, we briefly
review the main steps for convenient reference purposes.

Consider a Gaussian distribution for a stochastic variable $\ve x$,
\begin{eqnarray}
\pcal(\ve x) &=& \frac{\e^{-\frac12 (\ve y-\ma A\ve x)^T\ma C_n^{-1} (\ve y-\ma
    A\ve x)}}{\sqrt{|2\pi\ma C_n|}} \frac{\e^{-\frac12 \ve x^T\ma C_x^{-1} \ve
    x}}{\sqrt{|2\pi\ma C_x|}} \label{gengauss}
\end{eqnarray}
where $\ve y$ represents data (observations); $\ma A$ is some linear operator
that connects the unknown $\ve x$ to the data;
$\ma C_n$ is the noise covariance, which may or may not be diagonal;
and $\ma C_x$ is an optional covariance matrix for the prior of $\ve x$.
The value of $\ve x$ that maximizes the likelihood is readily found to be
\begin{equation}
\ve x_{\rm ML} = (\ma A^T \ma C_n^{-1}\ma A +\ma C_x^{-1})^{-1} \ma A^T \ma C_n^{-1} \ve y  \label{mlsolution}.
\end{equation}
Rather than finding the maximum-likelihood solution,
we now want to draw a random sample of the distribution in Eq.~(\ref{gengauss}).
This can be done by solving a linear system similar to that of Eq.~(\ref{mlsolution}),
but with a modified right-hand side $\ve b$, as
\begin{eqnarray}
\ve b &=&  \ma A^T \ma C_n^{-1} \ve y  
        +\ma A^T\ma C_n^{-1/2}\ve\omega_1 
        +\ma C_x^{-1/2}\ve\omega_2  \nonumber \\
\ve x  &=& (\ma A^T \ma C_n^{-1}\ma A +\ma C_x^{-1})^{-1} \ve b. \label{gibbssolution}
\end{eqnarray}
Here $\ve\omega_1$ and $\ve\omega_2$
are two vectors of random numbers drawn from a normalized 
Gaussian distribution $N(\ve 0,\tens I)$.
The first vector has length equal to that of the data stream $\ve y$,
the second one to that of $\ve x$.

We now want to apply this to our map-making problem.
We take turns in drawing samples from the
distributions of Eqs. (\ref{slice1}) and (\ref{slice2}).
We end up with the
following two-step procedure.
Starting from sample ($\ve m,\ve a$), we update the parameters as follows.
\begin{enumerate}
\item Map sampling (distribution of Eq. (\ref{slice1}), no prior):
\begin{equation}
\ve m' = (\ma P^T \ma C_w^{-1}\ma P)^{-1} [\ma P^T \ma C_w^{-1} (\ve y-\ma F\ve a) 
+ \ma C_w^{-1/2}\ve\omega_1] \label{gibbs1}
\end{equation} 

\item Noise sampling (Eq. (\ref{slice2})):
\begin{eqnarray}
\ve b &=&  \ma F^T \ma C_w^{-1} (\ve y -\ma P\ve m')
        +\ma F^T\ma C_w^{-1/2}\ve\omega_2 
        +\ma C_a^{-1/2}\ve\omega_3   \nonumber \\
\ve a' &=& (\ma F^T \ma C_w^{-1}\ma F +\ma C_a^{-1})^{-1} \ve b \label{gibbs2}
\end{eqnarray}
\end{enumerate}
To draw one pair of samples we thus need three independent Gaussian random vectors, 
$\ve \omega_1$, $\ve \omega_2$, and $\ve \omega_3$. Equations~(\ref{gibbs1}) and (\ref{gibbs2})
summarize our Gibbs map-making algorithm.

\subsection{Assumptions and optimization}

%We now proceed to apply the above procedure to simulated \Planck\ data.
The above discussion is general, and few assumptions regarding the
experiment in question are made. However, to make the formalism more
practical in terms of computer code implementation, we make a few
assumptions that are familiar from the conventional \Planck\ data
processing pipeline:
\begin{enumerate}
\item We ignore beam smearing effects, and assume that all signal
  recorded by a detector at a time comes from the pixel where the beam center falls.
  We are regarding the beam smoothing as a property of the sky itself,
  an assumption which is strictly valid only for symmetric beams.
  Exact treatment of an asymmetric beam requires full beam deconvolution,
  which is beyond the scope of this paper.
  With this assumption, the pointing matrix $\ma P$ is sparse, with three nonzero elements on each row. 
  corresponding to three Stokes components of one pixel. \\
\item We do not correct for bandpass effects, and assume that all detectors see the same sky. \\
\item Detector noise is uncorrelated between pointing periods. \\
\item Correlated noise within a pointing period is Gaussian and stationary.  \\
\item The white noise component, by definition, is uncorrelated from
  sample to sample, but not necessarily with constant variance. The
  covariance $\ma C_w$ is thus diagonal, but the values of the
  diagonal elements are allowed to vary. In the following we make the
  further simplifying assumption that the variance stays constant
  within one pointing period.  However, we allow for the possibility
  that some of the samples are discarded from the analysis by setting
  $\ma C_w^{-1}=\ma 0$.
\end{enumerate}

Under the these assumptions, the coupling matrix $\ma P^T\ma
C_w^{-1}\ma P$ is block-diagonal, and consists of a $3\times3$ block
per pixel that that may be easily inverted.  The map sampling step of
Eq.~(\ref{gibbs1}) represents a simple binning operation, where the
noise-cleaned TOI samples are coadded into pixels as given by the
pointing matrix $\ma P$, and the resulting map is normalized with the
corresponding $3\times3$ matrix.

The noise sampling step can be carried out independently for each
pointing period, which has an average length of about 40 minutes.
This leads to a significant memory saving when compared to full
map-making, and for the first time we can bring the baseline length
down to one single sample.  At this extreme limit $\ma F$ reduces to
the identity matrix, $\ma F=\ma I$, and the noise sampling step of
Eq.~(\ref{gibbs2}) simplifies to
\begin{equation}
\ve b =  \ma C_w^{-1} (\ve y -\ma P\ve m')
        +\ma C_w^{-1/2}\ve\omega_2 
        +\ma C_a^{-1/2}\ve\omega_3 \label{bdef}
\end{equation}
and
\begin{equation}
\ve a' = (\ma C_w^{-1} +\ma C_a^{-1})^{-1} \ve b. \label{filter1}
\end{equation}
Here the term $\ma C_a^{-1}\ve\omega_3$ represents the operation
of applying the noise filter to a sequence of Gaussian random numbers.
Since $\ma C_a$ is stationary, this can be carried out efficiently with FFT technique.

To increase the efficiency of the FFT operations, it is useful to pad
the data to a suitable FFT length.  In this
paper, we have chosen the following procedure to deal with data ends:
To reduce boundary effects, we first subtract the mean of the data
(over unflagged samples), and only then pad the data with zeroes to a
suitable FFT length. We then carry out the FFT operation; apply the
relevant filter; perform the inverse FFT; cut off the padded region;
and add back the original offset.

\subsection{Flagging and masking}

As mentioned regarding Eq.~(\ref{filter1}), $\ma C_a^{-1}$ represents
a stationary filter. However, the presence of missing observations
makes the full coupling matrix $\ma C_w^{-1} +\ma C_a^{-1}$
non-stationary. The two most common causes for missing observations
are glitches in the data collection, and the application of an
analysis mask that excludes bright foreground regions on the sky,
which is useful to prevent striping along the scanning path of the
instrument.

The most convenient way of implementing masking is formally by setting
the white noise level to infinity for the discarded samples, or,
equivalently, $\ma C_w^{-1}=\ma 0$.  The computational price for this is
that Eq.~(\ref{filter1}) can no longer be treated as a strict Fourier
filter.  The system can still be solved through conjugate gradient
iteration, but this is significantly slower than solving the
stationary system.  A faster, but approximate, solution is to fill the
gaps with a simulated data.  We consider both options more closely in
the following.

\subsubsection{Solving the non-stationary system}
\label{sec:nonstationary}

We first consider the exact solution of the non-stationary system in
Eq.~(\ref{filter1}), where $\ma C_w^{-1}=\ma 0$ in the flagged part of the
data.  Let us for convenience define
\begin{equation}
\ma A = \ma C_w^{-1} +\ma C_a^{-1},
\end{equation}
and $\ma M$ to be a stationary version of the same, with flags
appropriately omitted,
\begin{equation}
\ma M = \ma I/\sigma^2 +\ma C_a^{-1} .  \label{Mdef}
\end{equation}
If only a small fraction of samples are flagged, we have
\begin{equation}
{\ma M}^{-1} \ma A \approx \ma I.
\end{equation}

Conjugate gradient iteration can be sped up significantly with a
properly chosen preconditioner matrix. In our case, $\ma M$ represents
a natural choice for a preconditioner.  However, even with
preconditioning, full convergence typically requires 40--50 iteration
steps, which becomes computationally prohibitive when this algorithm
is repeated within every Gibbs loop.

Fortunately, we can speed up the procedure significantly by
reformulating the problem as follows.  We first symbolically write
the right-hand side of Eq.~(\ref{filter1}) in the form
\begin{equation}
 \ma A^{-1}\ve b = \ma M^{-1} {\ve b} + \mathrm{correction.}   \label{corr}
\end{equation}
Second, we define
\begin{equation}
\ma D=\ma I/\sigma^2-\ma C_w^{-1}
\end{equation}
to be the deviation from the constant white noise variance in $\ma
C_{w}$. Explicitly, $\ma D$ is a diagonal matrix with zero on the
diagonal for the non-flagged samples, and $1/\sigma^2$ for the flagged
ones. We denote by the number of flagged samples on the pointing
period by $N_{\rm f}$, and the total number of samples by $N_{\rm s}$.
Typically, only a small fraction of samples are flagged, and thus
$N_{\rm f}$ is much smaller than $N_{\rm s}$.  We then eigenmode
decompose $\ma D$ into $\ma D=\ma U\boldsymbol\updelta\ma U^T$, where now
$\boldsymbol\updelta$ is a diagonal matrix of size $(N_{\rm f},N_{\rm f})$ with
$1/\sigma^2$ on the diagonal. The matrix $\ma U$ has size $(N_{\rm
  s},N_{\rm f})$, and each column of $\ma U$ corresponds to one
flagged sample, and contains one nonzero (equal to one) element
marking its position on the data stream.

Using this notation we have
\begin{equation}
\ma A = \ma C_w^{-1} +\ma C_a^{-1} =
\ma M -\ma U\boldsymbol\updelta\ma U^T .
\end{equation}
Using the well-known Woodbury matrix identity
we can re-write the matrix inverse into the form
\begin{equation} 
\ma A^{-1} = \ma M^{-1} +\ma M^{-1}\ma U 
(\boldsymbol\updelta^{-1}-\ma U^T\ma M^{-1}\ma U)^{-1} 
\ma U^T\ma M^{-1}.
\end{equation}
The solution of Eq. (\ref{filter1}) therefore becomes
\begin{equation}
\ve a' = \ma M^{-1}\ve b +
\ma M^{-1}\ma U 
(\sigma^2\ma I-\ma U^T\ma M^{-1}\ma U)^{-1} 
\ma U^T\ma M^{-1}\ve b
\label{filter2}
\end{equation}
where we have written $\boldsymbol\updelta^{-1}=\sigma^2\ma I$.  This expression
conforms to the format of Eq.~(\ref{corr}), and is the solution we
were aiming at.

The vector $\ve a'$ can now be computed efficiently from
Eq.~(\ref{filter2}).  The first term, $\ma M^{-1}\ve b$, represents
the operation of applying the stationary filter of $\ma M$ to the
right-hand side $\ve b$, and can be carried out efficiently by FFT
techniques.  The rest represents a correction that takes care of
flagging.  Working from right to left, the matrix $\ma U^T$ represents
the trivial operation of picking the flagged samples from a full TOI,
and concatenating them into a vector of length $N_{\rm f}$.  The
middle matrix inversion must still be carried out using
conjugate gradient iteration, but the system is now much smaller than
the original one.  Finally, the matrix $\ma U$ represents the
operation of inserting the $N_{\rm f}$ values of its target vector
into a full-size TOI, in the positions of the flagged samples. 
We find that the solution of Eq.~(\ref{filter2}) requires only 5--6 iteration
steps for convergence, to be contrasted with 40--50 steps required for
the original system to converge.

\subsubsection{Gap filling}
\label{sec:gapfill}

As efficient as the procedure presented above is, solving the full
exact and non-stationary system still implies a computational cost
that is 5--6 times higher compared to a case where no flagging is
applied. In the following we therefore present an alternative way of
handling the flagged data sections: we fill them with a noise
realization, as part of the overall Gibbs sampling procedure. We note
that similar gap filling techniques have been routinely used by a long
history of previous CMB experiments, including \Planck\ LFI
\citep[e.g.,][]{planck2013-p02}, and we will in the following discuss
both exact and approximate solutions to this problem.

First, we introduce the white noise component $\ve w$ in the flagged
section as a new explicit Gibbs variable to be sampled over.  The
original noise sampling step,
\begin{equation}
 \ve{a'}  \leftarrow\pcal(\ve a \mid \ve{m};\ve y,\ma C_x),
\end{equation}
is therefore replaced by a two-step procedure
\begin{eqnarray}
 \ve{w'}  &\leftarrow& \pcal(\ve w \mid \ma C_w)   \nonumber \\
 \ve{a'}  &\leftarrow& \pcal(\ve a \mid \ve{m},\ve w;\ve y,\ma C_x).
\end{eqnarray}
The first step consists of generating a white noise sample for all
flagged TOI segments, which is simply equivalent to drawing a random
realization from the Gaussian distribution with variance $\ma
C_w=\sigma^2\ma I$.  The second step represents the usual sampling
operation of Eqs.~(\ref{bdef}) and (\ref{filter1}), but this time for a
full data stream where the gaps are filled, and consequently, the
white noise covariance $\ma C_w$ is uniform.

\begin{figure}
  \center
  \includegraphics[width=\linewidth]{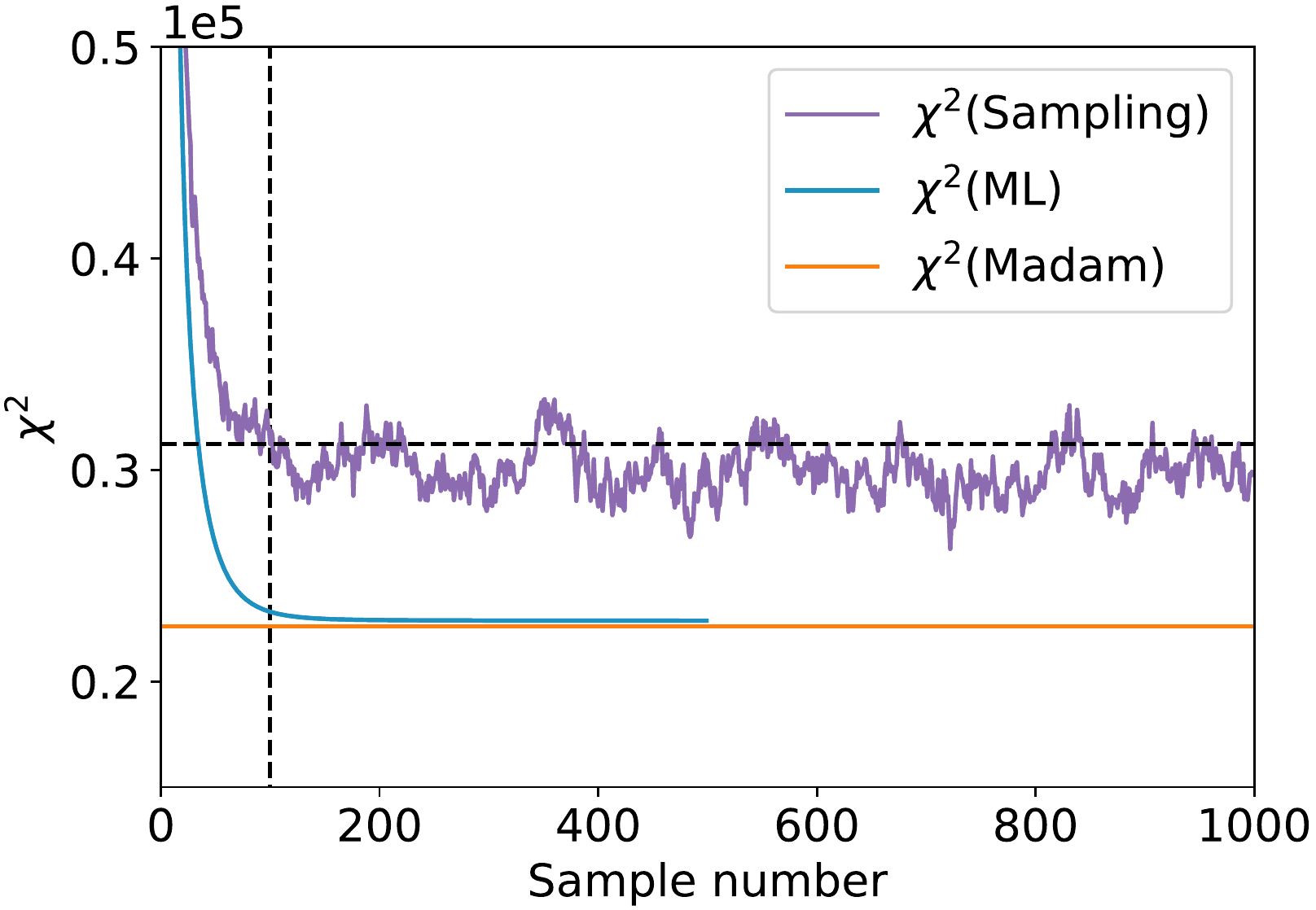}
  \caption{Gibbs chain burn-in as illustrated by the $\chi^2$
    difference between a given Gibbs sample and the noise-free 
    reference map. The dashed line shows the $\chi^2$ level after 100
    steps. The $\chi^2$ value from \Madam\ is shown by a horizontal line.}\label{fig:burn_in}
\end{figure}

\begin{figure}
  \center
  \includegraphics[width=\linewidth]{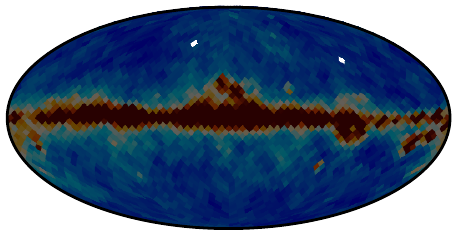}
	\caption{Location of \texttt{HEALPix} $N_{\textrm{side}}=16$ pixels 200 and
    600 on the sky in nested ordering.  The colour scale of the
    underlying map is dimmed to highlight the locations of the two
    pixels.}\label{fig:pixels}
\end{figure}

\begin{figure}
  \center
  \includegraphics[width=8.8cm]{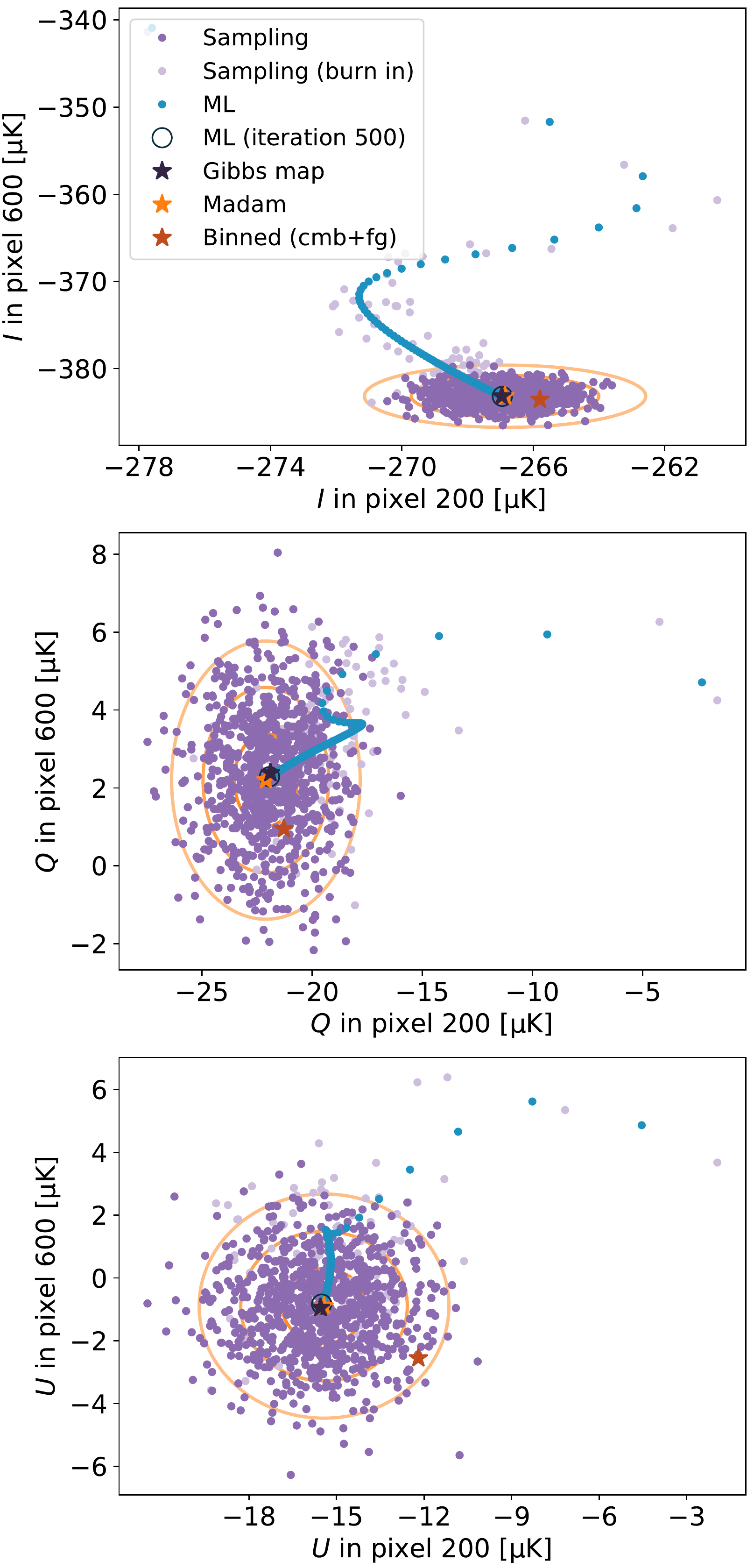}
  \caption{Gibbs chains for the Stokes parameters $I$, $Q$ and $U$ in
	pixels 200 and 600 in a \texttt{HEALPix}
    $N_{\textrm{side}}=16$ map with nested ordering. 
    The value in pixel 600 is shown on the $y$-axis, plotted against the value
    in pixel 200 ($x$-axis).
    Purple dots
    indicate individual steps during a sampling run, while blue dots
    indicate steps in a maximum-likelihood run. The orange stars
    indicate the \Madam\ solution, and orange contours indicate
    confidence regions based on white noise uncertainties alone.}\label{fig:chain}
\end{figure}

\begin{figure*}[t]
  \center
  \includegraphics[width=0.93\linewidth]{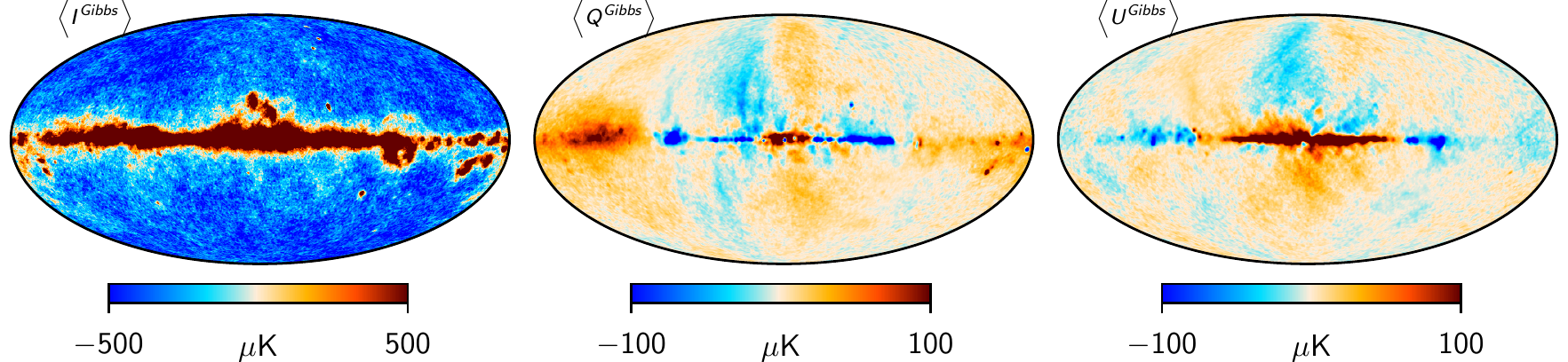}
  \includegraphics[width=0.93\linewidth]{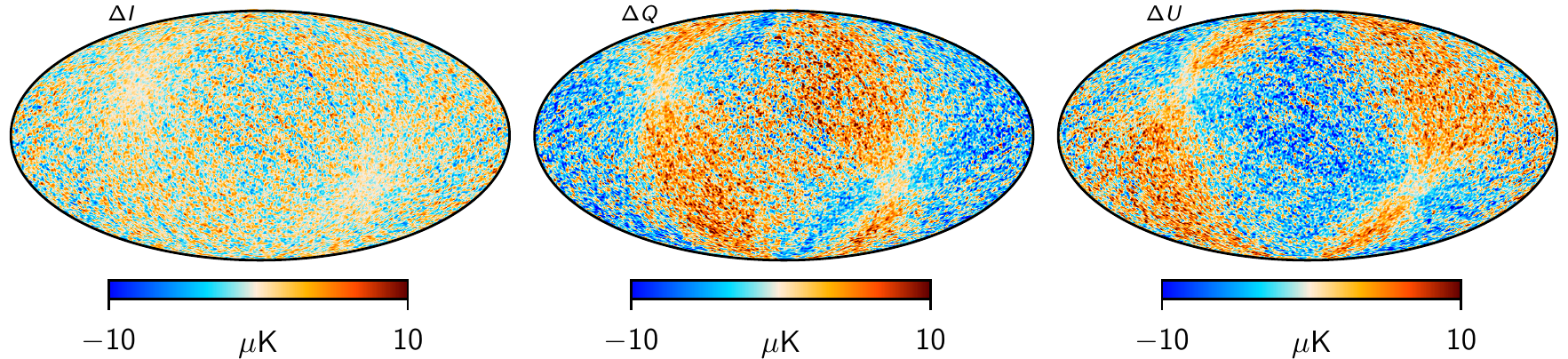}\\
  \includegraphics[width=0.93\linewidth]{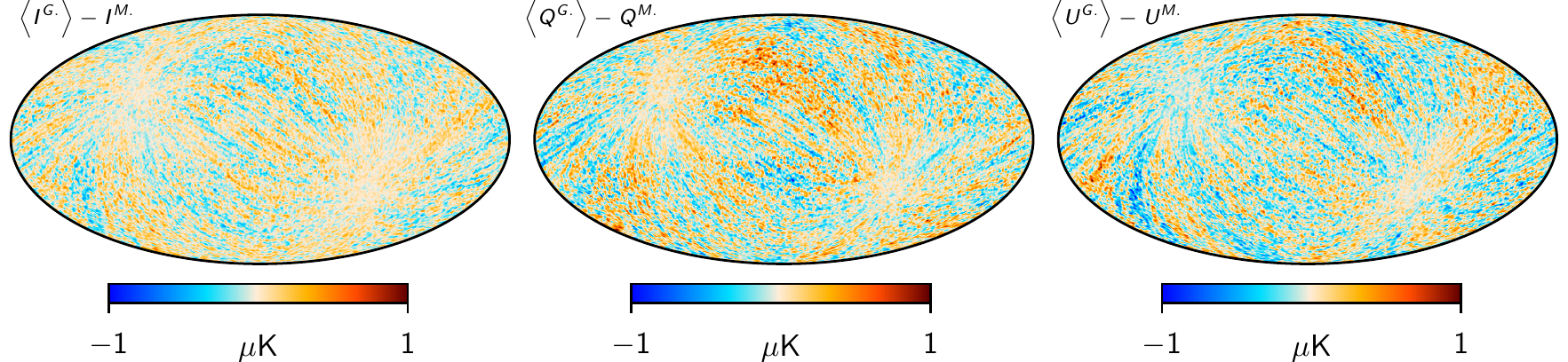}\\
    \includegraphics[width=0.93\linewidth]{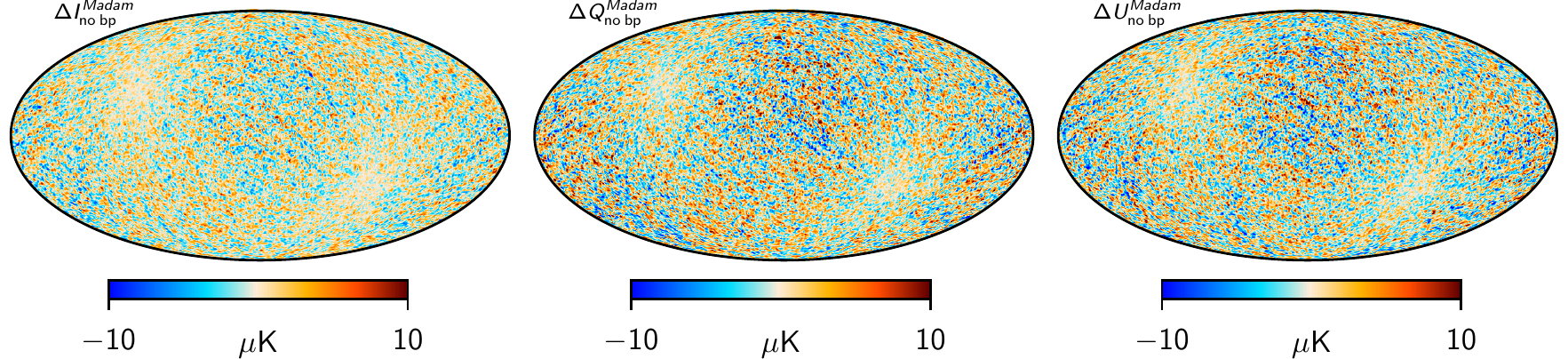}
  \includegraphics[width=0.93\linewidth]{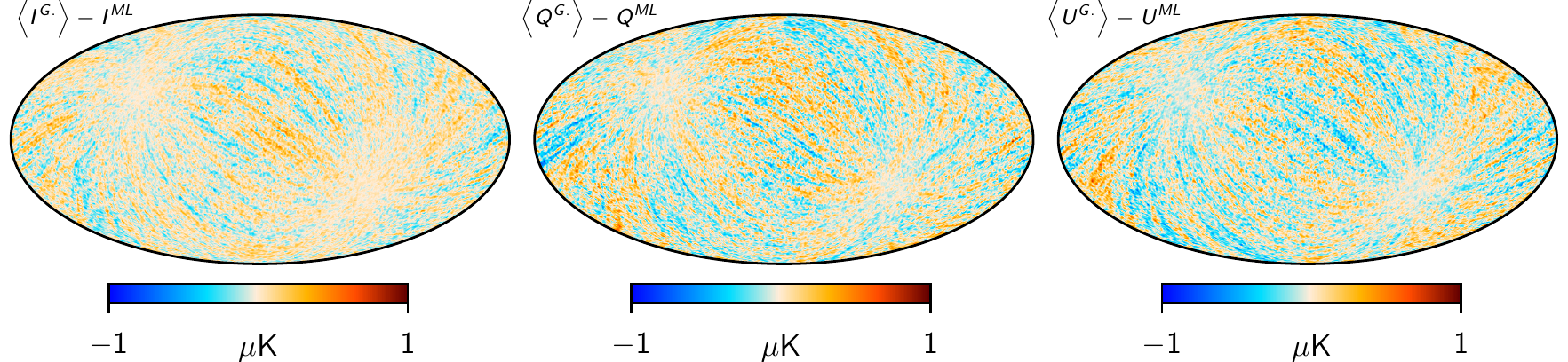}    
  \caption{Temperature and polarization maps from simulation. Columns show, from left to right, the Stokes $T$, $Q$, and $U$ components. 
    Rows show, from top to bottom, 
    1) Gibbs map: the mean of maps from the Gibbs chain, as evaluated from 900 samples; 
    2) absolute residual error as evaluated as the difference between the Gibbs map and the noise-free map; 
    3) the difference between the Gibbs map and the \Madam\ map derived from the same data set; 
    4) residual error in \Madam\ for a simulation that does not contain bandpass 
    (\emph{simulation that has same input foreground map for all radiometers}) evaluated as the difference 
    between the \Madam\ map and noise-free map; and 
    5) the
    difference between the Gibbs map and the maximum-likelihood map.  }\label{fig:gibbs_map}
\end{figure*}

The remaining question is, what should go in place of $\ve y-\ma P\ve m'$ in
Eq.~(\ref{bdef})?  This vector represents the current best estimate of
the noise stream, including both correlated and white noise.  In the
non-flagged data section, it is therefore raw data minus the sky
estimate, as usual.  In the flagged section, where no information is available
from the data regarding $\ve y$, it is given by the correlated noise
TOI, $\ve a$, from the preceding Gibbs iteration plus the white noise
realization $\ve w$ just generated, With the gaps filled in this manner, the calculation of
Eq.~(\ref{filter1}) can be solved with FFT technique without CG
iteration, and the overall algorithm therefore becomes very fast.

This solution has two potential disadvantages compared to the
original solution, worth keeping in mind. The first is a longer overall Monte Carlo
correlation length, which arises from the fact that we are now
conditionally sampling the correlated and white noise terms separately, which is
always less efficient than sampling them jointly. The second
disadvantage is the fact that this method increases the overall memory
requirements, since the correlated noise TOI for the flagged section 
also have to be stored in memory between two consecutive Gibbs iterations.
We test both solutions with simulations.

\section{Simulations}
\label{sec:simulations}

We characterize and validate the algorithms described above using
controlled simulated data.  Specifically, we create a set of simulated
time-ordered data using the {\tt{Level-S}} simulation
software \citep{reinecke2006}. We choose to focus on the \Planck\ LFI
30\,GHz channel, since the data set for this particular channel is the
smallest of the \Planck\ channels in terms of data volume, and
therefore also has the fastest turnaround time.  The simulated data
span the full \Planck\ LFI mission, i.e., four years of observations.

The simulated data contain time streams for CMB, foregrounds, and
noise. We simulate each component separately, and use the sum of all
time streams as an input to the Gibbs map-maker. The CMB signal is
simulated based on theoretical angular power spectra, while the
foreground model is adopted from a preliminary \BP\ analysis, taking
into account radiometer-specific beam and bandpass responses. Both CMB
and foreground simulations contain real \Planck\ main and intermediate
beams, as described by \citet{planck2014-a05}. We use {\tt{conviqt}}
\citep{prezeau2010} and {\tt{multimod}} tools from the
\Planck\ {\tt{Level-S}} package to convolve the sky with the beam, and to
scan a time-ordered data stream from the input sky according to the
detector pointing.  The noise TOI is simulated based on realistic
noise parameters, as described by \citet{planck2014-a03}, 
and generated with {\tt{multimod}}.

The \Planck\ LFI detector pointing is regenerated with \texttt{Level-S}, and
has been confirmed to agree well with the real \Planck\ LFI pointing.
Additionally, we use actual \Planck\ LFI flags to discard flagged
samples, and a processing mask described by \citet{bp13} to mask out
regions with strong gradients.

In addition we generate a set of 101 noise realizations by creating a
noise TOD with \texttt{Level-S} and calculating corresponding maps with
\Madam. We use this set of maps to study potential noise biases. The
noise realization used in the main validation work of the Gibbs
sampler is the first noise MC sample. Hence we have a total of 102
noise realizations.

\begin{figure*}
  \center
  \includegraphics[width=18cm]{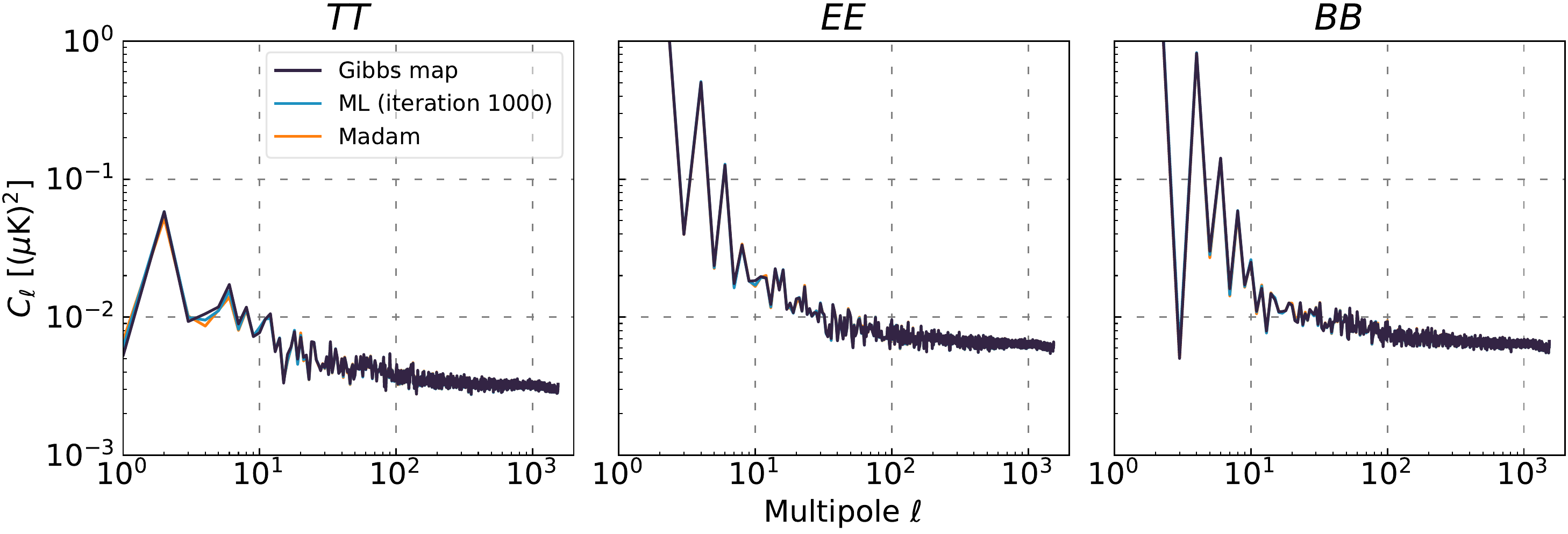}
  \caption{Residual noise spectra for different map-making options. 
  The residual noise spectrum is computed as the angular power spectrum of the residual noise map, which in turn is obtained
   as the difference between a map estimate and the noise-free map.}\label{fig:resnoise_comparison}
\end{figure*}

\begin{figure*}
  \center
  \includegraphics[width=18cm]{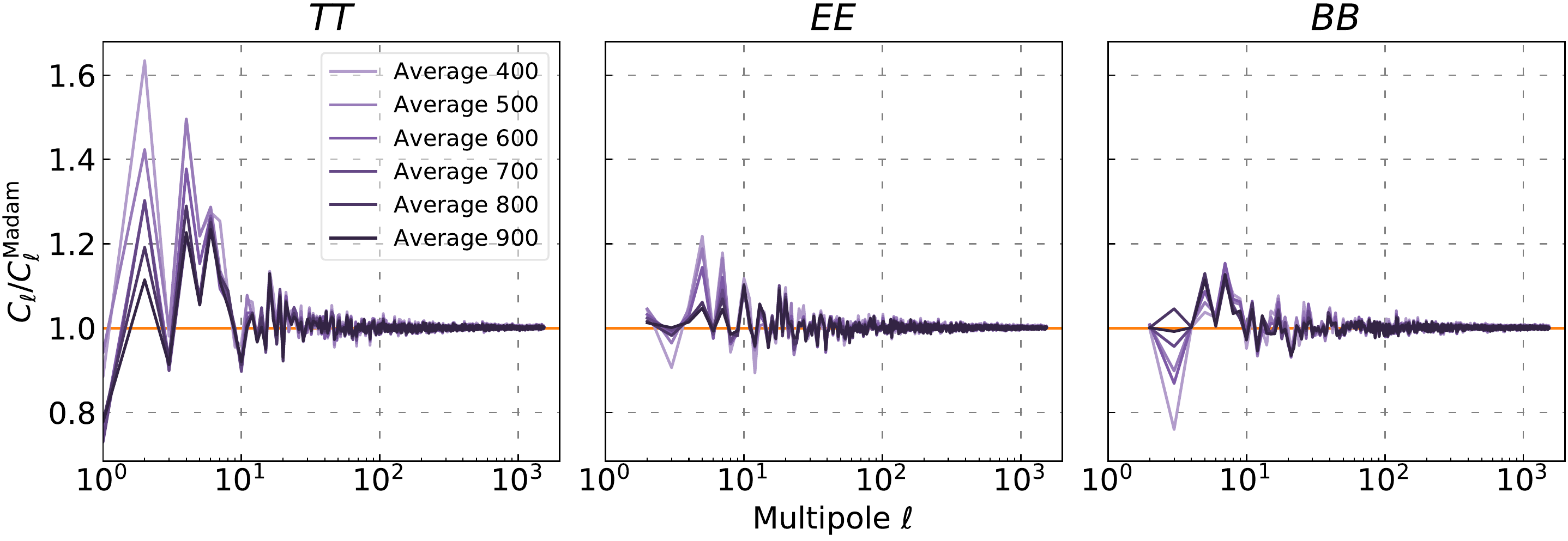}
  \caption{Ratio of the residual noise spectrum of the Gibbs map to that of \Madam, for different chain lengths.  }\label{fig:resnoise_SamplingOn}
\end{figure*}

\begin{figure*}
  \center
  \includegraphics[width=18cm]{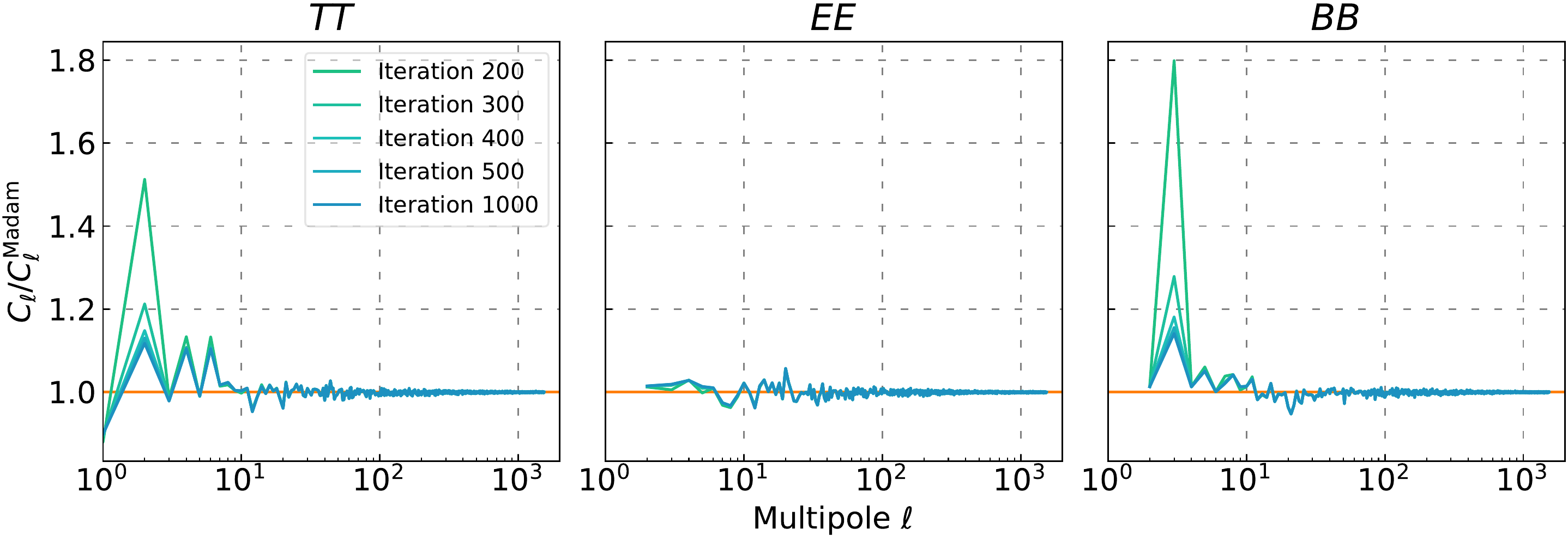}
  \caption{Ratio of the residual noise spectrum of the maximum-likelihood map to that of \Madam, for different Gibbs chain lengths.}
  \label{fig:resnoise_SamplingOff}
\end{figure*}

\begin{figure*}
  \center
  \includegraphics[width=18cm]{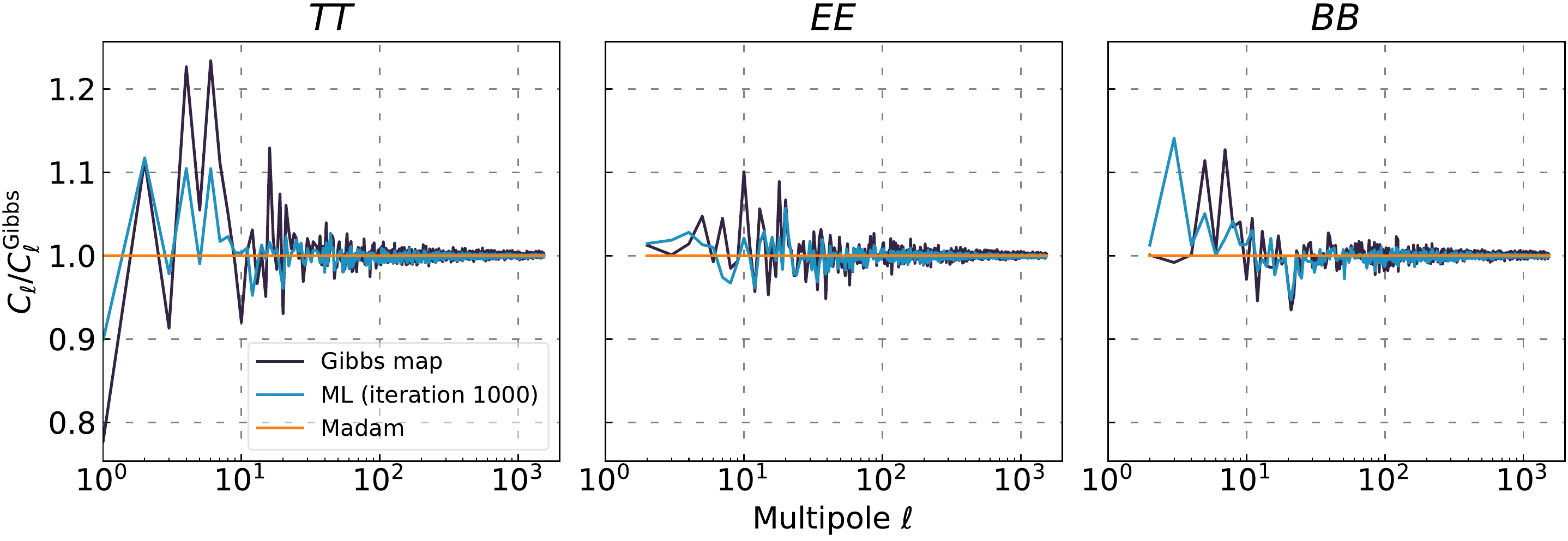}
  \caption{Ratio of residual noise spectrum to that of \Madam, for a Gibbs map and for a maximum-likelihood map
  with the largest available number of samples.
  The data are the same as in Figs. (\ref{fig:resnoise_SamplingOn}) and (\ref{fig:resnoise_SamplingOff}).
  }\label{fig:resnoise_spectra}
\end{figure*}

\begin{figure*}
  \center
  \includegraphics[width=18cm]{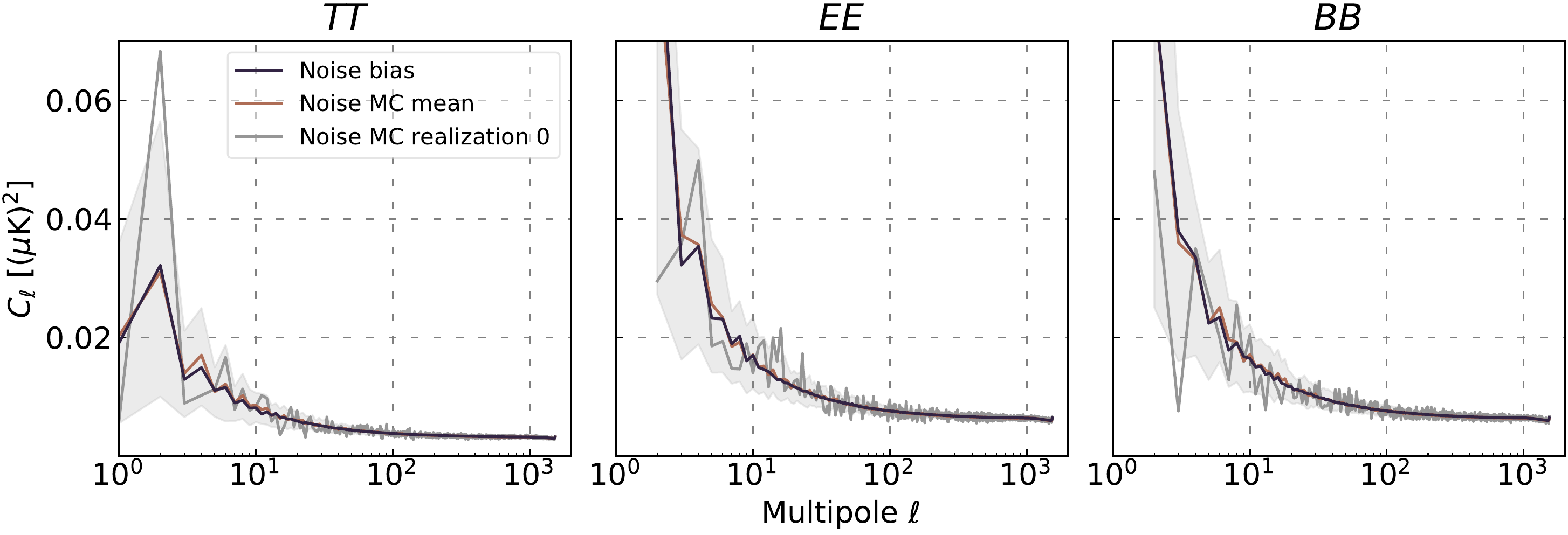}
  \caption{Noise bias from Gibbs chain compared with noise Monte Carlo. Noise MC contains 101 realizations, processed with \Madam. 
  The mean of the MC spectra is plotted in brown, and $\pm1\sigma$ regions in light grey. 
  Realization 0 of noise MC is plotted in medium grey.}\label{fig:gibbs_noisebias}
\end{figure*}

\begin{figure*}
  \center
  \includegraphics[width=18cm]{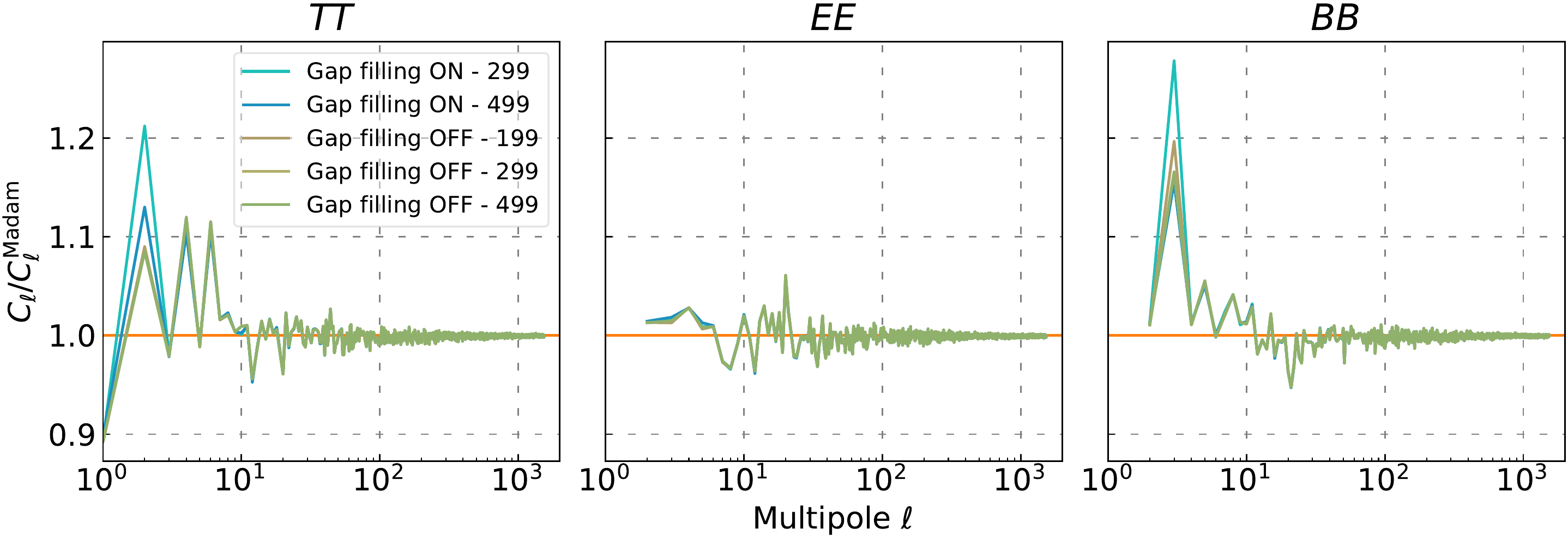}
  \caption{Ratio of residual noise spectrum to that of \Madam, with and without gap filling,
  for 199, 299, or 499 samples.}\label{fig:resnoise_gap}
\end{figure*}

\section{Results}

We now run the Gibbs sampling map-making algorithm described in
Sect.~\ref{sec:method} on the simulation data described in
Sect.~\ref{sec:simulations}.  As already mentioned, we use a test code
for this purpose that is independent of the \commander\ code that is
employed for the full \BP\ analysis, as summarized by \citet{bp01}.
This allows us to examine the map-making problem in isolation, and
also to try out different approaches with a faster turn-around time.
The test code implements the Gibbs sampling map-making procedure, as
presented in Sect.~\ref{sec:method}, without any component separation
step. The output maps thus contain both the CMB and foreground signal,
and represent traditional frequency maps, rather than astrophysical
component maps.

We run the code both in ``sampling mode'' and in ``maximum-likelihood
mode''.  In the sampling mode, the code performs full Gibbs sampling,
and draws samples from the combined baseline-plus-map posterior
distribution.  In the maximum-likelihood mode, the fluctuation terms
that involve $\ve\omega_1$, $\ve\omega_2$, and $\ve\omega_3$ in
Eqs.~(\ref{gibbs1}) and (\ref{gibbs2}) are omitted, with the effect that the
algorithm performs a (computationally very inefficient)
steepest-descent search towards the maximum-likelihood solution.

In the following, we refer to the posterior mean map, as evaluated
from the mean of the ensemble of posterior samples, as the ``Gibbs
map'', with a burn-in period appropriately excluded.  Thus, the Gibbs
map represents a single-point estimate of the true map.  In contrast,
the ``maximum-likelihood map'' refers to the last sample of a chain with
the sampling terms disabled.

Unless stated otherwise, we fill the gaps as described in
Sect.~\ref{sec:gapfill}, using the previous correlated noise sample as
baseline. In the noise sampling step, we mask the Galactic region to
reduce leakage of strong foreground  signal into polarization.
The mask is applied only in the noise sampling step, while in the
map-binning step, all data are included, such that a full-sky map is
produced as output.
We briefly discuss the other option, exact solution of the non-stationary system,
in section \ref{sec:cost}.

Because we are working with simulated data, we know exactly what the
true sky signal should be.  As a reference, we therefore construct a
map from the pure CMB-plus-foreground time stream, and exclude both
correlated and white noise.  We refer to this ideal map as the ``noise-free
map'', which represents the map we would obtain in the absence of
instrumental noise.  The difference between the Gibbs estimates and
this map provides a measure of residual noise in the former.

We construct another validation reference by running the
\Madam\ map-making code on the same simulated data.  \Madam, and the
new Gibbs sampler when run in the maximum-likelihood mode, solve
essentially the same likelihood problem, though through a different
implementation.  We thus expect the results to be very close, if not
identical.  Small differences may arise due different baseline
lengths, or from different implementation of the noise filter, or
simply from different convergence properties. In particular, since a
Gibbs sampler only allow changes along coordinate directions, we
expect this approach to converge much slower towards the
maximum-likelihood solution than a conjugate gradient solver like
\Madam. We note, however, that the Gibbs sampling approach will
normally not be used in the maximum-likelihood mode, but rather as a
sampler within a more complete analysis framework.

\subsection{Pixel space}

We begin by inspecting the maps visually.
Figure~\ref{fig:burn_in} shows the unnormalized $\chi^2$ difference,
computed as the squared difference between the noise-free map and a map from the Gibbs chain, as a function of sample number.
For each map we first subtract the mean of $I$ to remove arbitrary map offsets. 
This applies for all maps in this paper's analysis.
We interpret the first 100 steps as the ``burn-in'' period, and discard them from all further analysis.
In the maximum-likelihood mode, the $\chi^2$ value converges towards the \Madam\ value,
shown by a horizontal line.  In the sampling mode, the $\chi^2$ value includes the additional variation from 
the sampling.

We now pick two sky pixels for closer examination.  
We arbitrarily select two pixels at resolution $N_\mathrm{side}=16$
from the region excluding strong foregrounds, with pixel indices 200
and 600 in the \texttt{HEALPix} nested pixelization scheme; the locations of
these pixels are indicated in Fig.~\ref{fig:pixels},
where they are shown on top of the simulated CMB+foreground map.

In Fig.~\ref{fig:chain} we plot the $I,Q,U$ Stokes components of pixel 600 against those of pixel 200.
This creates a two-dimensional plot that illustrates the behaviour of the Gibbs sampling chain.
The correct signal value, as taken from the noise-free map, is indicated by a red star.  
The estimated values necessarily differ from this due to noise.
The \Madam\ estimate, representing the maximum-likelihood solution, is shown by an orange star.
As expected, the Gibbs sampler in the maximum-likelihood mode approaches the same solution, 
although we note that the number of steps required to reach it is large.

In the sampling mode, the map samples fluctuate around the maximum-likelihood solution,
revealing the shape of the likelihood distribution.
This can be compared to the theoretical white noise distribution.
Along with the maximum-likelihood solution for the map, \Madam\ produces a white noise
approximation of the noise covariance. This is indicated by orange circles in the same plot.
The Gibbs chain tracks the shape of the distribution well, when the burn-in period (shown by gray dots) is discarded.
The sampled distribution includes also a contribution from correlated noise,
which is missing in the white noise estimate.
As the residual is dominated by white noise, however,
these are difficult to discern in the pixel plot.
In Sect.~\ref{sec:harmonicspace} we will further examine the noise residuals in harmonic space,
where correlated residuals are visible much more clearly.

The Stokes components ($I,Q,U$) of the pixelized Gibbs map are shown
in the uppermost panel of Fig.~\ref{fig:gibbs_map}, adopting a \texttt{HEALPix}
resolution of $N_{\rm side}=512$, or $7\arcm$ pixel size.  All
polarization components in this figure have been smoothed with a $1\deg$
FWHM Gaussian smoothing window to reduce noise.

The second row of Fig.~\ref{fig:gibbs_map} shows the difference with
respect to the noise-free map. This represents residual noise and systematic errors 
in the Gibbs map.
We observe that, in addition to
white noise, there is also a large-scale signal residual in the
polarization component, indicating leakage of temperature signal into
polarization.  However, when we take the difference between the Gibbs
map and the \Madam\ map constructed from the same data, the structure
nearly disappears, as shown in the third row.  This indicates that the
same signal residual is present in the \Madam\ map as well.  The
leakage is thus not a property of the Gibbs sampling method, but a
feature common all map-making methods.

As shown below, the source of the leakage is detector-dependent
bandpass and beam mismatch, coupled to a frequency-dependent
foreground signal.  To mitigate this, we have masked the Galactic region
during correlated noise estimation. However, the mask size is a
trade-off: removing more of the foreground will reduce the leakage,
but at the same time will also remove useful data and make noise reduction more
difficult. Here, we are working with the 30 GHz channel, which is
particularly cumbersome in this respect, as the foreground signal
extends over a large part of the sky.

To pinpoint the origin of the large-scale residual in the second row,
we create a simplified simulation data set where the bandpasses of all
radiometers are set equal.  To do this, we took the bandpass-weighted input
sky for radiometer LFI27M, and fed it as input to all radiometers,
however always convolving with each radiometer's own beam.
Since running a full Gibbs chain is
computationally expensive, and we know that \Madam\ is subject to same
leakage, we only run \Madam\ on the new simulation.  The difference
between the \Madam\ map and the noise-free reference map is shown in
fourth row of Fig.~\ref{fig:gibbs_map}.  The structure is strongly
reduced as compared to the one seen in the second row.  This shows
that the large-scale structure seen in row 2 is caused by foreground
signal residuals in combination with differential instrument
responses, and associated with bandpass mismatch.  This highlights the
importance of combined map-making and systematics corrections, as
discussed by \citet{bp01}, and accounting for these types of effects
is one of the main motivations for the full \BP\ Gibbs sampler.

Finally, in the bottom row of Fig.~\ref{fig:gibbs_map} we show the
difference between the Gibbs map (with sampling), and the
maximum-likelihood map (last sample from the Gibbs chain with sampllng
off).  The differences between the two are below the 1~$\mu$K level.

\subsection{Harmonic space}
\label{sec:harmonicspace}

We now proceed to examine the noise residuals in harmonic space,
where the correlated residuals are seen more clearly.
In Fig.~\ref{fig:resnoise_comparison} we plot the residual noise spectrum
for \Madam, maximum-likelihood  map, and the Gibbs map.
The residual noise spectrum is computed as the angular power spectrum
of the difference map between a map estimate and the noise-free map.
All three spectra are very close to each other.

To show the differences more clearly, we plot ratios of these spectra
in Figs.~\ref{fig:resnoise_SamplingOn} and
\ref{fig:resnoise_SamplingOff}.  We use the \Madam\ spectrum as a
reference, and plot the ratio of other spectra to it.  In
Fig.~\ref{fig:resnoise_SamplingOn} we show the residual noise in the
Gibbs map, for different numbers of sampling steps.  In each case, the
map is an average of the Gibbs samples over the indicated number of
samples.  The burn-in period is excluded in all cases.  The spectrum
is very close to that of \Madam\ at high multipoles, but differences
are seen at the lowest multipoles. In particular, we see that the
Gibbs map converges very slowly with number of iterations,
resulting in an overall high computational cost. Clearly, a direct solver
 is highly preferred over the Gibbs approach in
applications that only require a single maximum-likelihood solution,
and no error estimate, as for instance is the case in most forward
simulation-based analysis pipelines.

One could have hoped for the Gibbs map to yield a lower noise
residual, due to the shorter baseline length. However, we do not
observe this in our simulations.  This is good news for the official
\Planck\ analysis, since it indicates that correlated noise at scales
below the \Madam\ baseline length (0.25 s) is already negligible.  For
a few multipoles ($\ell=4-6$) \Madam\ yields a lower residual than the
Gibbs solver.  This can be attributed to numerical differences in the
implementation of the noise filter, and is well below the sample
variance.

In a similar manner, Fig.~\ref{fig:resnoise_SamplingOff} shows the
ratio of residual noise spectrum in the maximum-likelihood map to that of
the \Madam\ map.  Here, the map in question is the last sample of a Gibbs chain
with the given number of steps, with sampling turned off. We see that
more than 500 steps are required for convergence. In
Fig.~\ref{fig:resnoise_spectra} we show in the same plot the 
residual spectra from Figs.~\ref{fig:resnoise_SamplingOn} and
\ref{fig:resnoise_SamplingOff}.  From both plots, we pick the result
with highest number of samples.

For the Gibbs sampling approach, the scatter of the samples around the
mean estimate provides a direct account of the residual noise, and may
be used to construct an estimate of the noise bias as follows:
We first compute individual noise maps by subtracting the average map
from each map sample.  We then compute the angular power spectrum of
each noise map with the standard {\tt anafast} tool
\citep{gorski2005}, and obtain the noise bias as the mean of these
spectra, dividing by $N-1$ as usual for a variance estimate.

In Fig.~\ref{fig:gibbs_noisebias} we plot this estimated noise bias
together with the average noise spectrum from 101 noise-only Monte Carlo samples,
as processed through \Madam.  The first MC sample is shown in dark
grey, and represents the same data set as the one used in the
validation of the Gibbs sampler.  The noise bias and the Monte Carlo
average agree very well.  We see also that the differences seen in
Fig.~\ref{fig:resnoise_spectra} at low multipoles are small compared
with the random scatter between realizations in the noise MC.
 
Note that unlike the residual noise spectra that require information
on the simulation input, the noise bias can be computed from Gibbs
chains alone, since it involves only the Gibbs samples and their mean.
The procedure is thus available when processing actual measurement
data.

\subsection {Computational costs}
\label{sec:cost}

We run the simulation on a 24-core compute node with Intel Xeon
E5-2697 (2.7 GHz) processors.  On this system, one \Madam\ run on the
simulation data set takes 4000 seconds of wall time (27
core-hours).  One Gibbs sampling step with gap filling takes 173
seconds, while the full cost of a Gibbs chain with 1000 steps is 1150
core-hours (2 days wall-clock time), or more than 40 times the cost of
a Madam run.  In maximum-likelihood the process is somewhat faster,
taking 107 second per iteration, and a total of 360 core-hours for 500
iteration steps; this is still 13 times the cost of \Madam.  It is thus
evident that if one is interested only in the maximum-likelihood
solution, the Gibbs procedure is not competitive with more direct
methods, such as \Madam.

We have also tried running Gibbs sampling without gap filling, which
is the most time-consuming step in the procedure, as it requires a
conjugate gradient iteration inside every Gibbs step.  Since we are
dealing here with a full-scale simulation with foreground signal and
bandpass effects, the convergence is somewhat slower than what we
observed with the pure noise simulation of
Sect.~\ref{sec:nonstationary}.  The procedure takes 1814 seconds per
Gibbs iteration, and is thus more than ten times slower than the
procedure with gap filling,
In our test case, the additional memory requirement associated with
the gap filling procedure is negligible in comparison with the
overall memory requirement. 

 In Fig.~\ref{fig:resnoise_gap} we compare the residual noise as
estimated both with gap filling and with the conjugate gradient
solution, both in the maximum-likelihood mode.  The conjugate gradient
option does converge faster, but the differences are in the end
negligible.

The most computationally expensive combination is the sampling mode
without gap filling.  Preliminary tests indicate that this option
would take two weeks of wall-clock time, most likely without offering
any benefit over the options already studied.  Thus we do not examine
this combination further.

\section{Conclusions}

We have introduced Gibbs sampling as a novel solution to the
map-making problem for CMB experiments.  Gibbs map-making can be run
in ``maximum-likelihood mode'' to find the map solution that maximizes
the likelihood function, or in ``sampling mode'' in which case the
algorithm draw samples from the corresponding posterior
distribution. The mean of this distribution gives a best estimate of
the underlying true sky, while the scatter of the samples provides an
account of the residual noise in the map estimate.

To validate the procedure, we have compared the results against those
of a direct maximum-likelihood solver (\Madam).  We have demonstrated
that with noise sampling disabled, the Gibbs sampling procedure can be
used to find the maximum-likelihood map solution.  Since the sampler
proceeds in small one-dimensional steps through the large
multi-dimensional parameter space, the procedure is relatively slow,
and offers little benefit over direct methods.  Even so, this is an
important validation step.  

The true value of the Gibbs sampling approach lies in error
propagation and integration with systematic corrections.  In this
paper, we have demonstrated the first part, namely that Gibbs sampling
provides an efficient method to sample the full likelihood
distribution surrounding the maximum-likelihood map.  This way we
obtain a reliable estimate the level of residual noise in the map
(noise bias), which is otherwise possible only through the
construction of a pixel-to-pixel noise covariance matrix, which is notorious as a
computationally demanding task, or through Monte Carlo
simulations. 

We have studied two ways to deal with gaps in the data: either solving
an exact maximum-likelihood system where flagged samples are assigned
an infinite variance, or including the gaps in the Gibbs process as
another variable to sample. The latter translates into filling the
gaps with a new white noise realization at every step.
In our test case, gap filling turns out to be significantly more efficient, and is thus our
 method of choice. It is worth bearing in mind, however, that gap filling comes with
 an increased memory requirement, which may become prohibitive in other user cases.

The full power of Gibbs sampling as a map-making solution is realized
in cases where map-making and noise removal are included as a part of
a full Gibbs sampling machinery, including foreground modeling and
estimation of instrument effects. This is the goal of the larger 
\BP\ project.

\begin{acknowledgements}
  We thank Prof.\ Pedro Ferreira for useful suggestions, comments and
  discussions, and Dr.\ Diana Mjaschkova-Pascual for administrative
  support. We also thank the entire \Planck\ and \WMAP\ teams for
  invaluable support and discussions, and for their dedicated efforts
  through several decades without which this work would not be
  possible. The current work has received funding from the European
  Union’s Horizon 2020 research and innovation programme under grant
  agreement numbers 776282 (COMPET-4; \BP), 772253 (ERC;
  \textsc{bits2cosmology}), and 819478 (ERC; \textsc{Cosmoglobe}). In
  addition, the collaboration acknowledges support from ESA; ASI and
  INAF (Italy); NASA and DoE (USA); Tekes, Academy of Finland (grant
   no.\ 295113), CSC, and Magnus Ehrnrooth foundation (Finland); RCN
  (Norway; grant nos.\ 263011, 274990); and PRACE (EU).
\end{acknowledgements}

\bibliographystyle{aa} 

\bibliography{../common/Planck_bib,../common/BP_bibliography}

\end{document}